\documentclass[12pt]{iopart}

\usepackage{iopams}

\usepackage{graphicx}
\usepackage{dcolumn}
\usepackage{float}
\usepackage{iopams}

\begin{document}

\title{\boldmath Evolution of electron-boson spectral density in the underdoped region of Bi$_2$Sr$_{2-x}$La$_x$CuO$_6$ \unboldmath}

\author{Jungseek Hwang$^{1}$ and J. P. Carbotte$^{2,3}$}
\address{$^1$Department of Physics, Sungkyunkwan University, Suwon, Gyeonggi-do
440-746, Republic of Korea\\ $^2$Department of Physics and Astronomy, McMaster University, Hamilton, Ontario L8S 4M1 Canada\\ $^3$The Canadian Institute for Advanced Research, Toronto, ON M5G 1Z8 Canada}

\ead{jungseek@skku.edu}

\date{\today}

\begin{abstract}

We use a maximum entropy technique to obtain the electron-boson spectral density from optical scattering rate data across the underdoped region of the Bi$_2$Sr$_{2-x}$La$_x$CuO$_6$ (Bi-2201) phase diagram. Our method involves a generalization of previous work which explicitly include finite temperature and the opening of a pseudogap which modifies the electronic structure. We find that the mass enhancement factor $\lambda$ associated with the electron-boson spectral density increases monotonically with reduced doping and closer proximity to the Mott antiferromagnetic insulating state. This observation is consistent with increased coupling to the spin fluctuations. At the same time the system has reduced metallicity because of increased pseudogap effects which we model with a reduced effective density of states around the Fermi energy with the range of the modifications in energy set by the pseudogap scale.

\end{abstract}
\pacs{74.25.Gz,74.20.Mn}

\maketitle

\section{Introduction}

Our understanding of the interactions that most importantly determine the low energy properties of the high critical temperature superconducting copper oxides remains incomplete and a challenge to the field. Structures on a scale of the order of phonon and/or spin fluctuation energies that are seen in various properties have found interpretation in terms of a phenomenological boson exchange model. In such an approach an electron-boson spectral density plays a central role and an important aim is to determine its magnitude and frequency dependence\cite{carbotte:2011}. In principle there will be a different electron-boson spectral function for each electron involved in the inelastic scattering and so it would also depend on initial state momentum. On the other hand in optics which is the case of interest here, all electrons are involved and the most relevant function is momentum averaged. For Raman scattering, the relevant average depends on the polarization of the incident light. For B$_{2g}$ (B$_{1g}$) the average involved is weighted more in nodal (antinodal) direction in the CuO$_2$ Brillouin zone. Of course anisotropies are known to play a role even in conventional metals\cite{leung:1976}. By contrast, angular resolved photoemission spectroscopy (ARPES) involves no momentum average but it is also a surface sensitive technique. Yet another average determines the critical temperature. In the gap channel the d-wave projection of the momentum discriminating spectral density would appear while in the renormalization channel it would be its average as in the normal state optical conductivity $\sigma(T,\omega)$ as a function of photon energy $\omega$ at temperature $T$.

Mathematical techniques have been developed\cite{schachinger:2006,heumen:2009} which make it possible to extract from optical data, an estimate of the underlying electron-boson spectral density $I^2\chi(\Omega)$ involved in the inelastic scattering of the charge carriers. The optical conductivity is related to microscopic quantities through the Kubo formula\cite{carbotte:1995} and this involves a two-particle propagator. The resulting expression for $\sigma(T,\omega)$ can be usefully rewritten as a generalized Drude form in terms of a memory function. The memory function plays a similar role for the conductivity as does the quasiparticle self-energy for the single particle propagator that enters ARPES. In direct analogy, an optical scattering rate $1/\tau^{op}(T,\omega)$ can be defined and taken as the input data for the inversion process whereby $I^2\chi(\Omega)$ is extracted through the use of a maximum entropy technique. To make this possible it was necessary to obtain a simplified but still quite accurate linear integral relationship between $1/\tau^{op}(T,\omega)$ and the unknown $I^2\chi(\Omega)$. Allen\cite{allen:1971} first obtained the explicit form for the kernel in this integral equation. It applied at $T =$ 0 and was generalized to finite $T$ by Shulga {\it et al.}\cite{shulga:1991}. Both derivations assumed that the band structure density of states could be taken constant over the energy range of interest. When this no longer applies because of pseudogap formation as is the case in the underdoped region of the high $T_c$ cuprates, Sharapov and Carbotte\cite{sharapov:2005} provided a new formula valid at any temperature $T$ which features in addition to $I^2\chi(\Omega)$, the effective self consistent electronic density of states $\tilde{N}(\omega)$. Their expression reduced, as it must, to the earlier formula of Mitrovic and Fiorucci\cite{mitrovic:1985} when the zero temperature $T \rightarrow$ 0 limit is taken. Energy dependence in the electronic density of states can strongly affect properties of both normal and superconducting state\cite{schachinger:1990}.

In this paper we employ a maximum entropy technique\cite{schachinger:2006,jaynes:1957} to obtain information on the electron-boson spectral density $I^2\chi(\Omega)$ in the underdoped region of Bi$_2$Sr$_{2-x}$La$_x$CuO$_6$ (Bi-2201) from the recently published optical data in this cuprate family by Dai {\it et al.}\cite{dai:2012}. An important aim is to investigate how $I^2\chi(\Omega)$ evolves as the doping ($p$) is reduced and the Mott antiferromagnetic insulating state is approached. Most of the previous work\cite{schachinger:2006,heumen:2009}, but not all\cite{hwang:2012}, has not included effects of pseudogap formation\cite{hufner:2008} in the inversion process. Here we also generalized the technique to handle finite temperature data.

\section{Theoretical frame work}

\begin{figure}
\vspace*{-1.4cm}%
\centerline{\includegraphics[width=3.5in]{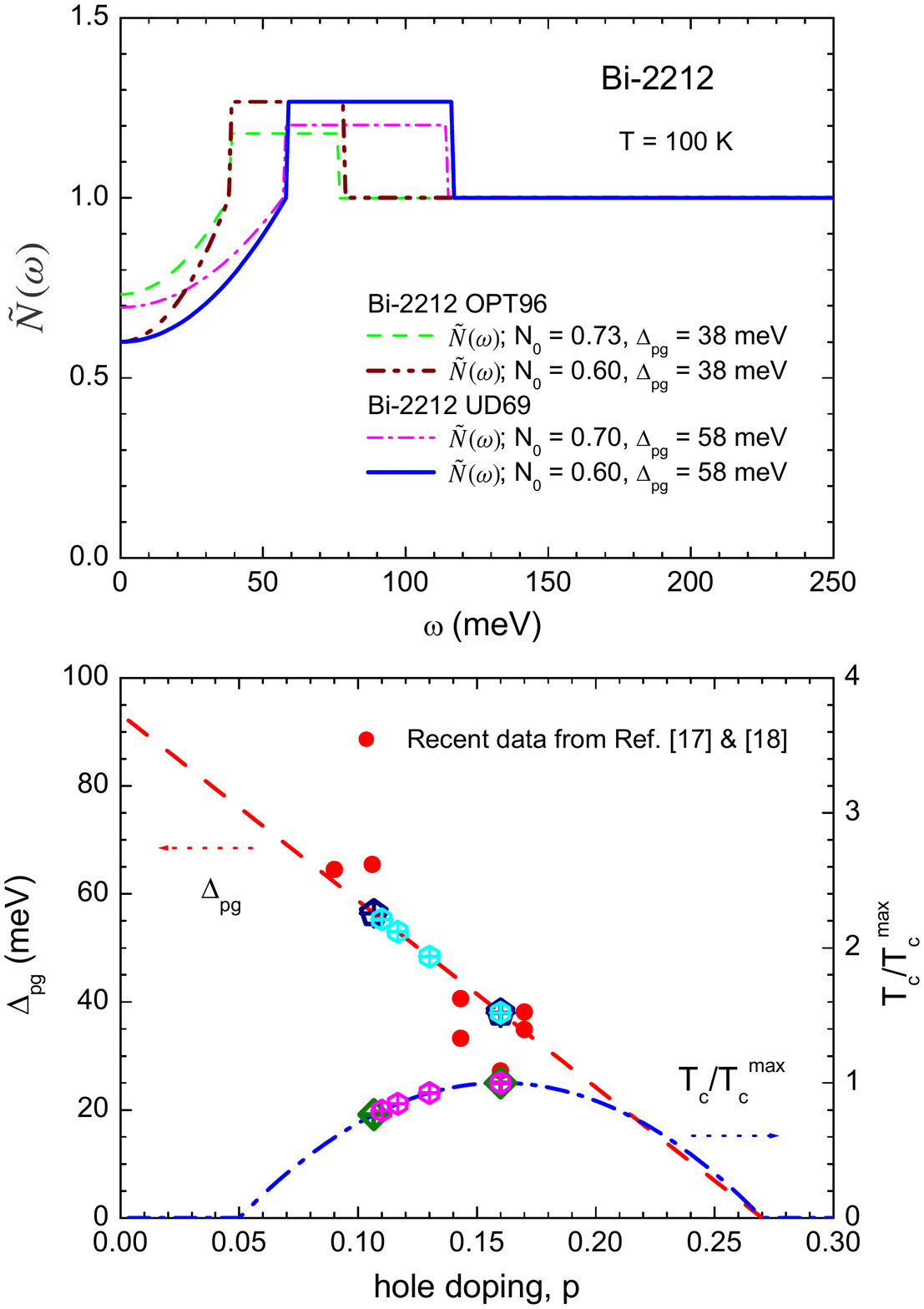}}
\vspace*{-1.0cm}%
\caption{(Color online) Top frame is the model used for the effective density of states $\tilde{N}(\omega)$ in the pseudogap phase. Parameters appropriate to Bi-2212 OPT96 with pseudogap $\Delta_{pg} =$ 38 meV and UD69 with $\Delta_{pg} =$ 58 meV are used. Two values of $\tilde{N}(\omega =0)$ (denoted by $N_0$) are considered $N_0 =$ 0.73 (dashed green) and 0.6 (dashed-double dotted brown) for OPT96 and 0.7 (dash-dotted magenta) and 0.6 (solid blue) curve. Bottom frame gives the phase diagram for Bi-2201 and Bi-2212. The lower dash-dotted (blue) curve gives $T_c/T_c^{max}$ versus doping $p$ (right hand scale). The straight dashed (red) line gives the value of the pseudogap size from the work of H\"{u}fner {\it et al.}\cite{hufner:2008} (left side scale in meV). The solid red circle symbols are from Ref. \cite{ma:2008} and \cite{kurosawa:2010}.}
\label{fig1}
\end{figure}

We begin with the generalized Drude formula for the temperature $T$ and photon energy $\omega$ dependent optical conductivity\cite{schachinger:2006}
\begin{equation}\label{eqn1}
\sigma(T,\omega) = \frac{i}{4 \pi}\frac{\Omega_p^2}{\omega-2\Sigma^{op}(T,\omega)}
\end{equation}
where $\Omega_p$ is the plasma frequency and $\Sigma^{op}(T,\omega)$ is the optical self-energy related to the memory function. It is conventional to write $-2Im\Sigma^{op}(T,\omega)$ as the optical scattering rate $1/\tau^{op}(T,\omega)$ which is $T$ and $\omega$ dependent. The real part $-2Re\Sigma^{op}(T,\omega)$ is written in terms of an optical effective mass $m^{* op}$ as $[m^{* op}(T,\omega)/m-1]\omega$ where again it is dependent of both energy and temperature. The two functions $\tau^{op}(T,\omega)$ and $m^{* op}(T,\omega)$ are not independent but are related by a Kramers-Kronig transform. Experimentalists often measure the reflectivity of their sample from which they obtain $\sigma(T,\omega)$. Many also provide the optical scattering rate which can be obtained from their knowledge of the conductivity. The relationship is
\begin{equation}\label{eqn2}
\frac{1}{\tau^{op}(T,\omega)} \equiv \frac{\Omega_p^2}{4 \pi}Re\Big{[} \frac{1}{\sigma(T,\omega)}\Big{]}.
\end{equation}
Here we start with the data for $1/\tau^{op}(T,\omega)$ given by Dai {\it et al.}\cite{dai:2012} in Bi-2201 and for comparison we will also use similar data by Hwang {\it et al.}\cite{hwang:2004} for two samples of Bi-2212 namely UD69 (underdoped with $T_c =$ 69 K) and OPT96 (optimally doped with $T_c =$ 96 K). The relationship between optical scattering rate and the electron-boson spectral density $I^2\chi(\Omega)$ which we wish to obtain is given by the Kubo formula which gives the optical conductivity in terms of the microscopic parameters of the materials of interest. In general a two-particle correlation function is involved with vertex corrections. Simplifications are needed if one is to make progress. For the case when the electronic density of states $\tilde{N}(\omega)$ is approximately constant over the energy range of interest in optical experiment P. B. Allen\cite{allen:1971} derived a very simple, yet as it has turned out, quite accurate formula that directly relates $1/\tau^{op}(\omega)$ to the electron-boson spectral density $I^2\chi(\Omega)$ which applies at $T =$ 0. He found
\begin{equation}\label{eqn3}
\frac{1}{\tau^{op}(\omega)} = \frac{2 \pi}{\omega}\int_{0}^{\omega}d\Omega I^2\chi(\Omega)(\omega-\Omega).
\end{equation}
Such a linear integral equation remains an ill-defined inversion problem. Nevertheless maximum entropy techniques can be applied to obtain an estimate of $I^2\chi(\Omega)$ from the information on $1/\tau^{op}(\omega)$\cite{schachinger:2006,hwang:2012}. Allen's derivation is based on lowest order perturbation theory (Fermi Golden Rule) with electron-boson spectral density $I^2\chi(\Omega)$ appropriate for transport. This function includes a weighting factor which emphasizes the enhanced effect of backward scattering in depleting current. This weighting is absent in the equilibrium spectral density which enters quasiparticle properties. Formula (\ref{eqn3}) can also be obtained directly from the Kubo formula for the dynamic conductivity in a boson exchange mechanism treated within an Eliashberg\cite{marsiglio:1998} formulation. In such an approach a weak coupling assumption is used to achieve the simplification involved in Eqn. (\ref{eqn3}) and vertex corrections are included at the level of changing the spectral density $I^2\chi(\Omega)$ from its equilibrium to its transport value. An important conclusion of that work is that formula (\ref{eqn3}) is quantitatively accurate. The same approach, based on the Kubo formula, was subsequently used by Sharapov and Carbotte\cite{sharapov:2005} to include both a pseudogap and finite temperature effects. In terms of the effective density of states $N(\omega)$, which now has an energy dependence, they found
\begin{eqnarray}\label{eqn4}
\frac{1}{\tau^{op}(\omega,T)}&=&\frac{\pi}{\omega}\int_{0}^{\infty}d\Omega I^2\chi(\Omega)\int_{-\infty}^{\infty}dz[N(z-\Omega)+N(-z+\Omega)] \\\nonumber
&\times& [n_B(\Omega)+1-f(z-\Omega)][f(z-\omega)-f(z+\omega)]
\end{eqnarray}
which is still linear in $I^2\chi(\Omega)$ and a maximum entropy inversion technique still applies but the Kernel is now more complicated and depends both on temperature through the bose $n_B(\Omega)$ and fermi $f(z \pm \Omega)$ thermal factors and on the self-consistent effective density of states $\tilde{N}(z)\equiv[N(z)+N(-z)]/2$. Note this quantity is symmetrized about the Fermi energy at $z =$ 0. In zero temperature limit Eqn. (\ref{eqn4}) was obtained previously by Mitrovic and Fiorucci\cite{mitrovic:1985} using a very different method. They generalized directly the Fermi Golden Rule approach of Allen to the case of an energy dependent density of electronic states. As discussed in Ref \cite{hwang:2012} Eqn. (\ref{eqn4}) at zero temperature can be used to invert data on optical scattering rate in the standard way for a given model of $\tilde{N}(z)$ which must first be specified. In principle this function knows about the mechanism by which the pseudogap forms. But here we model it through a fit to experimental data. In this paper we have generalized our maximum entropy inversion codes to include the thermal factors of Eqn. (\ref{eqn4}) as well as our model for the effective density of states. This allows us to invert normal state data at any temperature $T$. In Fig. \ref{fig1} (top frame) we show the model we have used for $\tilde{N}(z)$ in all our numerical works. $\tilde{N}(\omega)$ is taken to have a value of $N_0$ at $\omega =$ 0, it then increases as $\omega^2$ to values 1.0 at $\omega = \Delta_{pg}$ above the pseudogap the state lost below $\omega = \Delta_{pg}$ are distributed uniformly in the range from $\Delta_{pg}$ to twice $\Delta_{pg}$ after which $\tilde{N}(\omega)$ becomes 1.0 (i.e. no changes over its constant base band structure value). The $\tilde{N}(z)$ can be written as
\begin{eqnarray}\label{eqn4a}
\tilde{N}(z)\!\!&=&\!\!N_0\! +\! (1\!-\!N_0)\Big{(} \frac{z}{\Delta_{pg}}\Big{)}^2\:\:\mbox{for} \:|z|\leq \Delta_{pg} \nonumber \\
&=&\! 1+\frac{2(1\!-\!N_0)}{3} \:\:\:\:\:\:\:\mbox{for}\: |z|\in (\Delta_{pg},2\Delta_{pg}) \nonumber \\
&=& \!1 \:\:\:\:\:\:\:\:\:\:\:\:\:\:\:\:\:\:\:\:\:\:\:\mbox{for} \:\:|z|> 2\Delta_{pg}.
\end{eqnarray}
The parameters that can be varied are $N_0$ related to the depth of the pseudogap well and the value of $\Delta_{pg}$. The value of the pseudogap $\Delta_{pg}$ is known from the work of H\"{u}fner {\it et al.}\cite{hufner:2008}. These authors have surveyed the results of many experiments which give information on the size of pseudogap, $\Delta_{pg}$ and have concluded that it rises linearly with decreasing doping $p$, starting from zero at the upper side of the superconducting dome as shown (dashed red line) in the lower frame of Fig. \ref{fig1}. This is to be viewed as the behavior on average and a guide only. For the specific case of Bi-2201 not part of the H\"{u}fner {\it et al.}'s survey, an early scanning tunnelling microscopy (STM) study\cite{kugler:2001} gave a small value of the pseudogap of order 10 meV. More recent data however give much larger values. Ma {\it et al.}\cite{ma:2008} find a large gap of order $\sim$27 meV for optimal doping and $\sim$10 meV for highly overdoped samples. Kurosawa {\it et al.}\cite{kurosawa:2010} find a large gap of 30 meV for optimal and as large as 60 meV for highly underdoped ($p \sim$ 0.1). These values are not so different from those suggested in the work of H\"{u}fner {\it et al.}\cite{hufner:2008} shown in Fig. \ref{fig1} lower frame as the dashed red line. These values are also consistent with the size of the pseudogap temperature $T^*$ measured by NMR\cite{zheng:2005,kawasaki:2010} in Bi-2201 which are of the same order as measured in other high transition temperature cuprates\cite{timusk:1999} lending further support for the phase diagram we have used in this study (lower frame of Fig. \ref{fig1}).

With the value of $\Delta_{pg}$ given, a single parameter remains to be specified in our density of states model $\tilde{N}(\omega)$ namely $N_0$. In Ref. \cite{hwang:2012} we found that the inversion process was not very sensitive to the details of the density of state variation with energy but was mainly dependent on the number of lost states below $\omega = \Delta_{pg}$ which are recovered above this energy. Here we denote this quantity by $PG_{loss}$ and favor it as the single measure of reduced metallic behavior brought about by the opening of the pseudogap.

\section{Main results}

\begin{figure}
\vspace*{-1.4cm}%
\centerline{\includegraphics[width=3.5in]{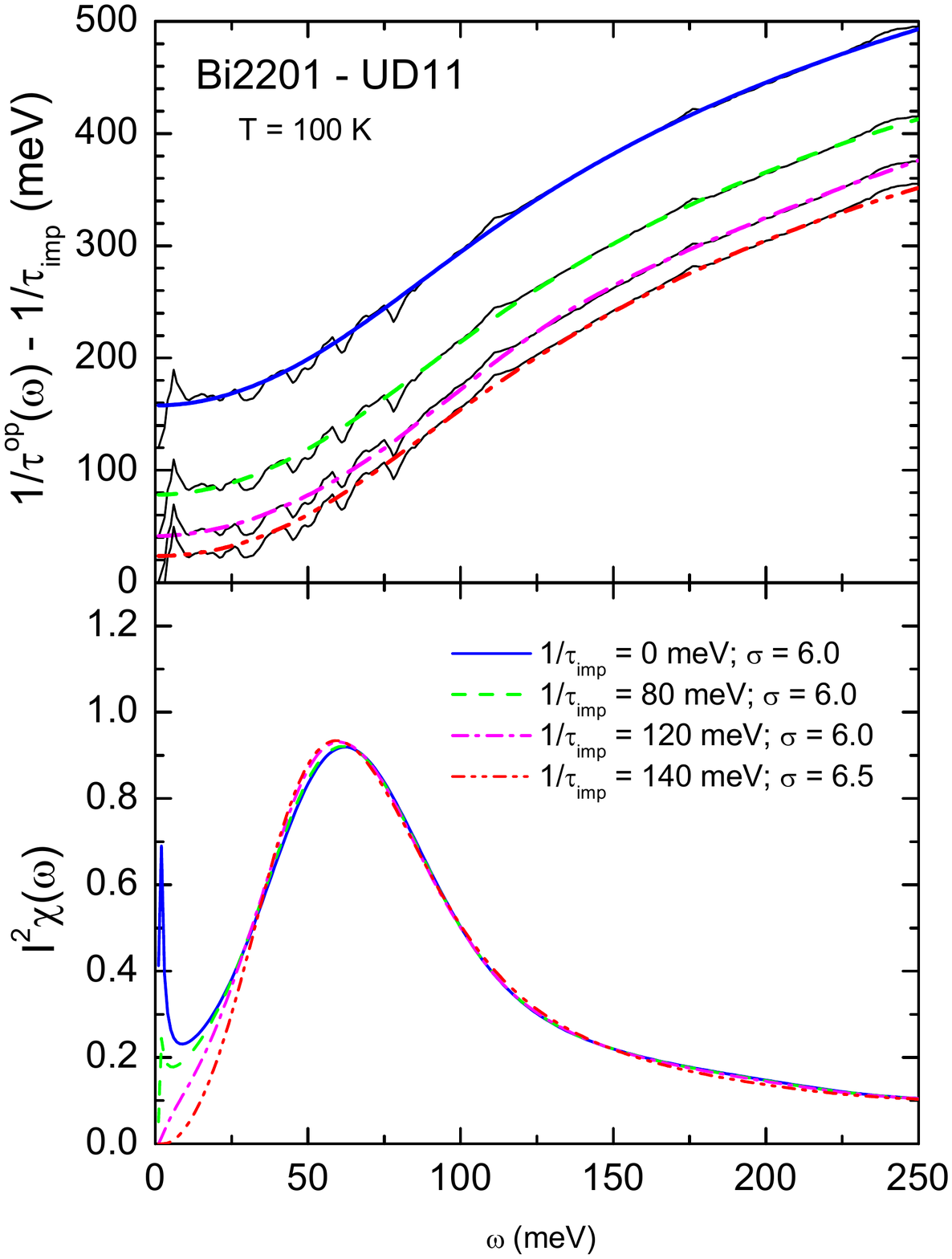}}
\vspace*{-1.0cm}%
\caption{(Color online) Top frame gives the optical scattering rate in Bi-2201: UD11 from the work of Dai {\it et al.}\cite{dai:2012}. The data is in the normal state at $T =$ 100 K (light solid black curves). We have subtracted from the raw data constant residual scattering rates, $1/\tau_{imp}$, of 0 meV (solid blue), 80 meV (dashed green), 120 meV (dash-dotted magenta) and 140 meV (dash double dotted red) from top to bottom. The fits to the data are as labeled. The bottom frame gives the resulting electron-boson spectral densities $I^2\chi(\omega)$ in the same notation as the curves in the upper frame. Note the sharp unphysical rises in the spectral density at very low $\omega$ for no impurity scattering and 80 meV subtracted cases.}
\label{fig2}
\end{figure}

First results are found in Fig. \ref{fig2} which applies to UD11 (underdoped with $T_c =$ 11 K) sample of Dai {\it et al.}\cite{dai:2012} at $T =$ 100 K. The data for the optical scattering rate was read off their figure (3b) and is reproduced in the top frame as the light solid black lines. For the top curve, no residual scattering was subtracted off the data while for other low curves we subtracted $1/\tau_{imp} =$ 80, 120, and 140 meV (as marked in the figure) before proceeding with the maximum entropy inversion. In all cases temperature was set to 100 K in the kernel of Eqn. (\ref{eqn4}). The solid blue curve is our fit to the top curve. Double dot dashed red curve is our fit to the data curves with $1/\tau_{imp} =$ 140 meV. Fits to $1/\tau_{imp} =$ 80 meV (dashed green) and 120 meV (dash-dotted magenta) are also shown. Except for the lowest curve the fits are equally good with ($\sigma =$ 6.0). Here $\sigma$ is related to the quality of the maximum entropy fit to the data, as explained in reference \cite{schachinger:2006}, the default model for $I^2\chi(\omega)$ is a constant.

Turning next to the bottom frame of Fig. \ref{fig2} we see that both solid blue and dashed green curves for $I^2\chi(\omega)$ show unphysical upturns as $\omega \rightarrow$ 0 which we trace to the fact that in both these cases we have not subtracted from the optical scattering rates enough of a residual part before inversion and the maximum entropy program is trying to accommodate residual scattering by including a $\omega =$ 0 contribution to $I^2\chi(\omega)$. We understand this as follows. In an electron-boson exchange formulation, impurity scattering can be simulate with a model $I^2\chi(\omega)\sim \omega \:\delta(\omega)$. In all cases from this point on we first subtract a $1/\tau_{imp}$ from the data of reference \cite{dai:2012} chosen so that the unphysical upturn in $I^2\chi(\omega)$ noted has just disappeared. Subtracting a larger elastic impurity scattering only changes the small $\omega$ behavior of the spectral density and has little consequences on many of the average properties of $I^2\chi(\omega)$ such as the area under it. But it does change the first inverse moment, which is related to the value of the electron-boson mass renormalization. The weighting factor $1/\omega$ enhances the effect of small $\omega$ in $I^2\chi(\omega)$ but this is not very important when the unphysical rise in the spectral density at small $\omega$ has been removed. In all cases we take the value of residual scattering rate $1/\tau_{imp}$ to be the value at which the unphysical divergence in $I^2\chi(\omega)$ as $\omega \rightarrow 0$ just disappears in our inversions. These values are given in Fig. 3 and 4.

\begin{figure}
\vspace*{-1.4cm}%
\centerline{\includegraphics[width=3.5in]{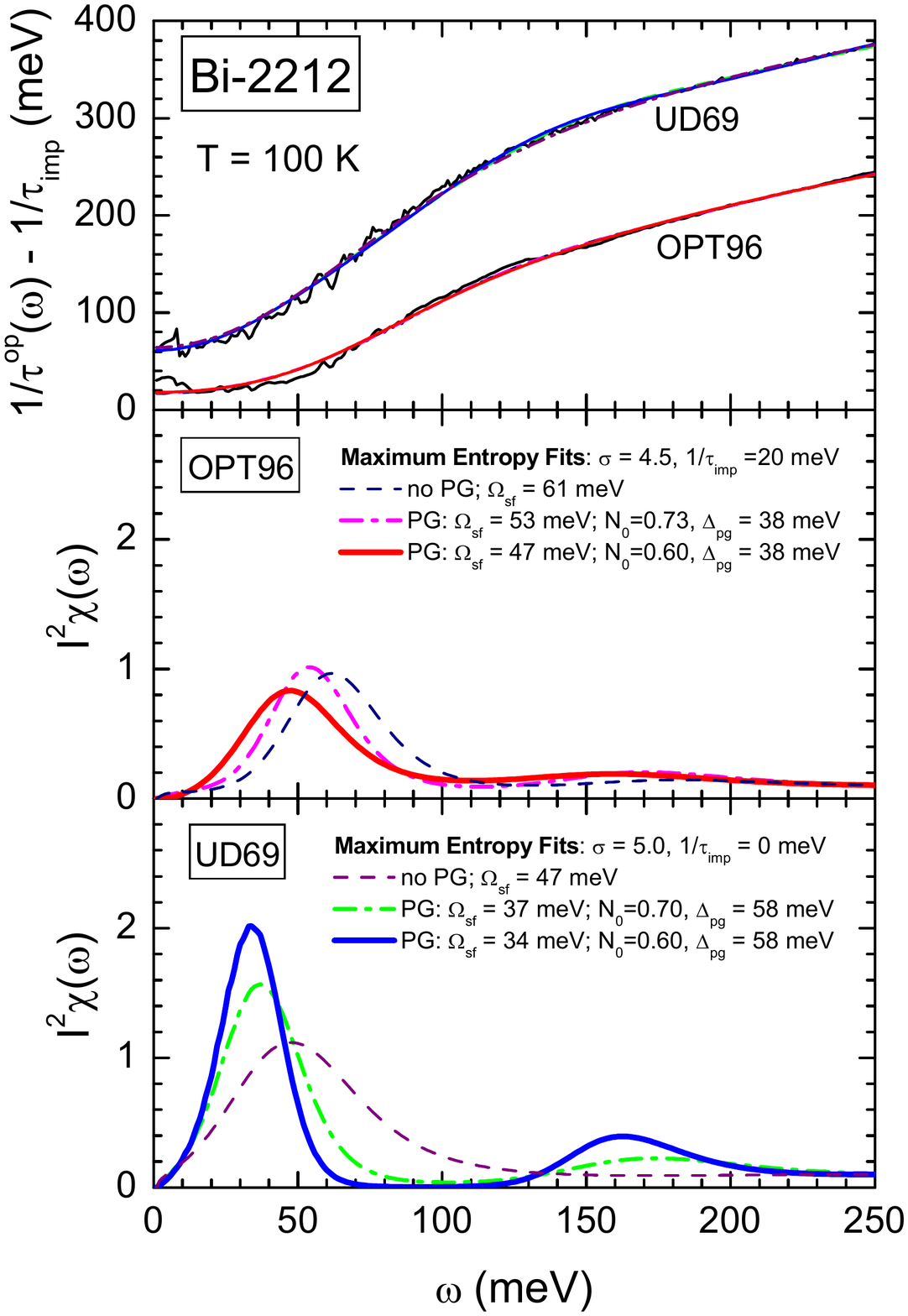}}
\vspace*{-1.0cm}%
\caption{(Color online) Top frame is the optical scattering rate (solid black curve) for Bi-2212 taken from the work of Hwang {\it et al.}\cite{hwang:2007a}. The upper curve is for UD69 while the lower curve is for OPT96 with a residual scattering $1/\tau_{imp}$ of 20 meV subtracted from the raw data. The dashed dark blue line is our maximum entropy fit assuming there is no pseudogap. The dash-dotted (pink) curve includes a pseudogap of $\Delta_{pg} =$ 38 meV and a well depth $N_0 =$ 0.73 while for the solid red curve $N_0 =$ 0.60. These are for OPT96 while for UD69, the purple dashed curve has no pseudogap, green dash-dotted has $\Delta_{pg} =$ 58 meV with $N_0 =$ 0.70 and solid blue has $N_0 =$ 0.60. The two lower frames show our results for the recovered electron-boson spectral density $I^2\chi(\omega)$. The line type and colors are the same as for the top frame. All spectra show a peak at $\Omega_{sf}$ and a background extending to high energies. The value $\Omega_{sf}$ that is obtained is smaller for the underdoped sample as compared with optimally doped. Including a pseudogap shifts $\Omega_{sf}$ to lower energies as does lowering the value of $N_0$. }
\label{fig3}
\end{figure}

\begin{figure}
\vspace*{-1.4cm}%
\centerline{\includegraphics[width=3.5in]{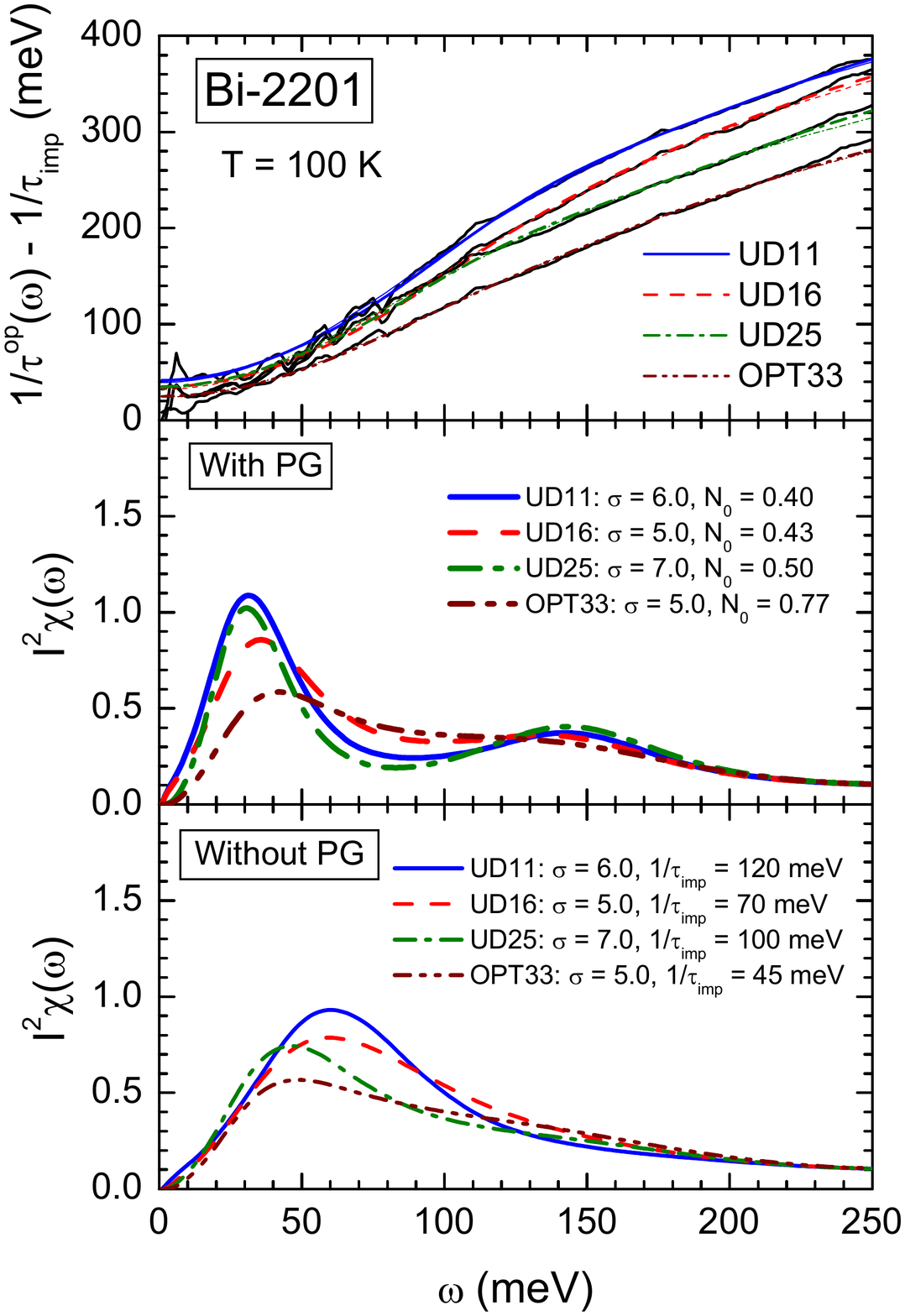}}
\vspace*{-1.0cm}%
\caption{(Color online)Top frame, the optical scattering rate (solid black curves) for Bi-2201 from the work of Dai {\it et al.}. The four samples with our maximum entropy fits are UD11 solid blue, UD16 dashed red, UD25 dash-dotted dark green and OPT33 dash-double dotted brown curves. The recovered electron-boson spectral densities are shown in the middle and bottom frames where the line types and colors are the same as for the top frame. The heavy curves (middle frame) include a pseudogap while the light curves (bottom frame) do not. The parameter $\sigma$ controls the quality of the fit with default model a constant (see Ref. \cite{jaynes:1957}), \cite{schachinger:2006} and \cite{hwang:2012}}
\label{fig4}
\end{figure}

\begin{figure}
\vspace*{-1.4cm}%
\centerline{\includegraphics[width=3.5in]{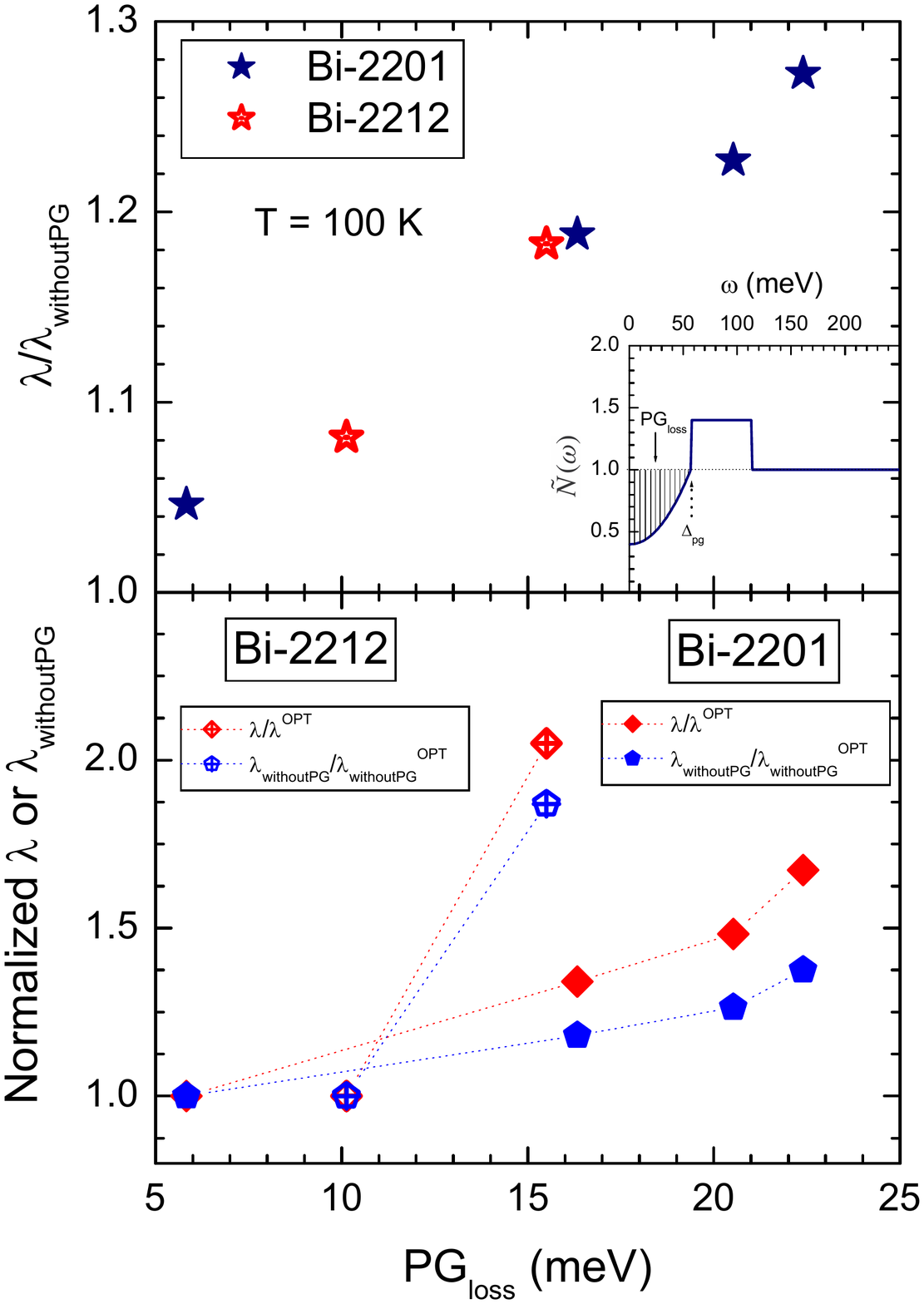}}
\vspace*{-1.0cm}%
\caption{(Color online) Top frame gives the ratio of the electron-boson mass renormalization factor ($\lambda$) including a pseudogap to its value without the pseudogap. The solid dark blue stars are for Bi-2201 and the empty red stars are for Bi-2212. For Bi-2212 with pseudogap you use $N_0 =$ 0.60 for both OPT96 and UD69 (see Fig. \ref{fig3}). On the horizontal axis we use as variable $PG_{loss}$ in meV, which measures the relative number of states below $\omega = \Delta_{pg}$ (shaded region in the inset) which are transferred to the region above $\Delta_{pg}$ in the effective density of states. The inset provides a sketch of this transfer of states. Bottom frame gives the ratio of $\lambda$ normalized to the value at optimum doping. The open red diamonds give twice the first inverse moment of $I^2\chi(\omega)$ when a pseudogap is included in our maximum entropy inversion and the blue open pentagons are the case when the pseudogap is not included. The solid symbols are for Bi-2201 in contrast to the open symbols for Bi-2212 which are included for comparison.}
\label{fig5}
\end{figure}

While we are mainly concerned in this paper with Bi-2201 data\cite{dai:2012}, it is nevertheless of interest to provide a comparison with the Bi-2212 family as it is closely related. In that instance, maximum entropy inversion of the optical data is already available but these where limited to the optimum plus overdoped regime of the phase diagram. This restriction was imposed on previous work because no pseudogap was included in the formalism available at that time. The top frame of Fig. \ref{fig3} shows our results for two samples at $T =$ 100 K namely OPT96 lower curve and UD69 upper curve. The solid black curve is data from reference\cite{hwang:2004,hwang:2007a}. The dashed dark blue and dash-dotted pink are our fits to the OPT96 sample, without ($N_0 =$ 1.0) and with pseudogap ($N_0 =$ 0.73). Here $\Delta_{pg}$ was taken to be 38 meV from reference\cite{hufner:2008}. In agreement with our previous finding the peak in the case $N_0 =$ 1.0, which is at 61 meV, has shifted down to 53 meV when a pseudogap is accounted for (middle frame). If $N_0$ is reduced to 0.60 the peak $\Omega_{sf}$, which is the spin-fluctuation scale, is shifted further to 47 meV (solid red curve). For the UD69 sample the pseudogap is larger, 58 meV (see Fig. \ref{fig1} bottom frame). Its effect on the recovered $I^2\chi(\omega)$ is shown in the bottom frame of Fig. \ref{fig3}. The solid blue curve is for $N_0 =$ 0.60, the dash-dotted green is for $N_0 =$ 0.70, and the dashed purple is for $N_0 =$ 1.0 (i.e. no pseudogap is accounted for). Note that the peak energy goes from 47 meV  ($N_0 =$ 1.0) to 37 meV and 34 meV with $N_0 \neq 1.0$. Including the pseudogap has a very significant effect on the recovered electron-boson spectral density and this needs to be accounted for. In principle the emergence of a pseudogap changes the electronic structure and hence should lead to corresponding modifications in the electron-boson exchange spectral density $I^2\chi(\omega)$. As our maximum entropy technique effectively determines this function through a fit to the optical data itself, these effects are included automatically in the recovered $I^2\chi(\omega)$ function.

In Fig. \ref{fig4} we show our results for the Bi-2201 series of reference\cite{dai:2012}. The data (solid black curve) was read off from their Fig. (3b). In all cases we are guided in our choice of pseudogap well depth at zero energy (denoted here by $N_0$) by the observation made in angular resolved photo emission (ARPES)\cite{kanigel:2006} that the data are consistent with a Fermi arc model with arc length proportional to the reduced temperature $t = T/T^*$ where $T^*$ is the pseudogap temperature given in Ref \cite{dai:2012}. As described by Hwang {\it et al.}\cite{hwang:2008b} this implies in our density of state model that $N_0$ also goes like $T/T^*$. This fixes this parameter i.e. $N_0$ = 100 K/$T^*$ where the pseudogap temperature $T^*$ is taken from the data of Ref \cite{dai:2012}. We have also verified in our numerical work that small deviation from this chosen value makes no qualitative changes to our recovered $I^2\chi(\omega)$ spectra. These $N_0$ values are given in the middle frame of the figure. The pseudogaps ($\Delta_{pg}$) used for fittings are 38, 49, 54, and 56 meV for OPT33, UD25, UD16 and UD11, respectively (see also the lower frame of Fig. \ref{fig1}). In accordance with our previous finding we subtracted a residual scattering contribution as noted in the figure before proceeding to the inversion. Our fits to the data on the optical scattering rates are shown in the top frame, the highest solid blue for UD11, dashed red for UD16, dashed-dotted dark green for UD25 and dash-double dotted brown for OPT33. The middle and bottom frames give our recovered electron-boson spectral densities with and without pseudogap, respectively. Trends are similar to those found in Fig. \ref{fig3}. Other optical data in Bi-2201, some with Pb doping, have appeared and have been inverted\cite{heumen:2009,heumen:2009a} to recover a boson spectral density. A histogram is used for $I^2\chi(\omega)$ with parameters determined through a least square fit to the conducting data. These works do not include a pseudogap and are based on an expression for the dynamic conductivity which is obtained under the assumption that the electronic density of states is constant over the entire band taken to be infinite. The results are in reasonable agreement with those shown in the lower frame of Fig. \ref{fig4} which were obtained taking $\Delta_{pg} =$ 0. In particular the position in energy of the main peak agree. In both analysis the mass enhancement factor $\lambda$ increases with decreasing doping $p$. The absolute value of $\lambda$ is larger in Ref. \cite{heumen:2009a} than what we find here. This may be related to the different treatment of the residual scattering.

In Fig. \ref{fig5} we give results for the mass renormalization factor $\lambda$ which has often been taken as the most important single measure of the associated renormalization. By definition
\begin{equation}\label{eqn5}
\lambda =2\int_0^{\infty}\frac{I^2\chi(\Omega)}{\Omega} \: d\Omega
\end{equation}
Because the $I^2\chi(\omega)$ used in this equation was extracted from optical conductivity data, it is a transport spectral density and includes vertex corrections. Although closely related, it is distinct from the equilibrium spectral density which enters ARPES data and distinct from that which enters Raman which has its own vertex. A comparison of these various electron-boson exchange spectral densities is given in Ref \cite{carbotte:2011} for the case of optimally doped Bi-2212. In each case the first inverse moment of $I^2\chi(\omega)$ which defines the single number $\lambda$ in Eqn. (\ref{eqn5}) provides a useful measure of the magnitude of the renormalizations involved. What is plotted in Fig. \ref{fig5} is the ratio $\lambda/\lambda_{withoutPG}$ which is seen to always be greater than one. On the horizontal axis we use the parameter $PG_{loss}$ which is defined in the inset. The area of the shaded region in meV defines $PG_{loss}$ and provides a measure of the relative number of electronic states that are transferred from the region $\omega \leq \Delta_{pg}$ to the region $\omega > \Delta_{pg}$ by the opening of the pseudogap. More explicitly $PG_{loss} = \frac{2}{3} \Delta_{pg} (1 - N_0)$. As the "metallicity" is reduced i.e. $PG_{loss}$ is increased, the mass enhancement parameter also increases monotonically over its no pseudogap value. We take this to mean that as the Mott insulating state is more closely approached by decreasing the doping $p$ toward zero (half filling) the coupling to the bosons is also increased. If we assume that the major coupling represented in $I^2\chi(\omega)$ is to spin fluctuation, this would make sense since the proximity to the antiferromagnetic state is also increased. In the bottom frame we provide additional information and compare in each case the increases in $\lambda$ over its value for optimum doping. For Bi-2212 the increase is a factor of 2 while for Bi-2201 it is smaller but still of order 50 \%. Even when no pseudogap is included in the inversion process, blue symbols, this large increase in $\lambda$, as we more closely approach the antiferromagnetic state, remains.

\section{Summary and conclusions}

 Structures or so called "kinks" observed\cite{lanzara:2001,damascelli:2003,borisenko:2006} in the dressed electronic dispersion curves on the high T$_c$ copper oxides have been widely interpreted as due to interaction with bosons. These structures are analyzed in terms of an electron-boson spectral density function $I^2\chi(\omega)$ which provides a phenomenological description of the low energy scattering processes\cite{carbotte:2011}. Such an approach is also at the basis of the nearly antiferromagnetic Fermi liquid model of Pines and coworkers\cite{millis:1990,monthoux:1994,chubukov:2002} which has enjoyed considerable success in understanding the measured properties of the cuprates\cite{chubukov:2002}. Some theoretical results based on the Hubbard and t-J models have also provided strong evidence that a boson exchange model does indeed emerge in strongly correlated systems with electron-boson spectral density related to the exchange of spin fluctuations\cite{maier:2008}. The available experimental literature has recently been reviewed\cite{carbotte:2011} and the conclusion made that quite consistent spectral densities are obtained using very different experimental techniques on the same material. Angular resolved photo emission\cite{schachinger:2008,bok:2010,zhang:2012}, scanning tunnelling microscopy\cite{levy:2008}, Raman scattering\cite{muschuler:2010} and dynamic optical conductivity data provide a frequency dependence for $I^2\chi(\omega)$ which all point to the same mechanism involving the exchange of spin fluctuations. Further, when comparison with inelastic spin-polarized neutron scattering data is available, agreement\cite{hwang:2007,hwang:2008c,yang:2009} between the shape of $I^2\chi(\omega)$ and that of the local spin susceptibility $\chi(\omega)$ is found. A particularly noteworthy case is for La$_{2-x}$Sr$_{x}$CuO$_4$ where the agreement with the detail spin susceptibility data of Vignolle {\it et al.}\cite{vignolle:2007} is remarkable\cite{hwang:2008c}. It has also been noted\cite{carbotte:2011} that the scale difference found between the optically derived spectral density and its ARPES counterpart is due to vertex corrections present in transport but absent in equilibrium properties such as ARPES. The reader is referred to Ref \cite{carbotte:2011} for more details. These facts support for the concept of a boson exchange mechanism as a useful phenomenology that helps researchers better understand the nature of the effective correlation in these materials. It also allows one to correlate in a simple frame work, many of their observed properties.

 Here we used a maximum entropy technique to extract from optical data an estimate of the electron-boson spectral density which, for the first time, includes both finite temperature effects and a model pseudogap. The pseudogap enters the formula for the optical scattering rate only through the effective density of states symmetrized about the Fermi energy $\tilde{N}(\omega)$. To model $\tilde{N}(\omega)$ we employ the known linear increase of $\Delta_{pg}$ with decreasing doping as well as an informed mathematical form for the $\omega$ dependence of $\tilde{N}(\omega)$. This leaves a single unknown parameter, the depth of the pseudogap well or more precisely its value at $\omega =$ 0. i.e. $N_{0}$. We set its value through consideration of the length of the Fermi arc measured by ARPES\cite{kanigel:2006,hwang:2008b} in the underdoped cuprates. Alternatively one can take the relative amount of spectral weight in the electronic density of states which is transferred from the energy region below $\Delta_{pg}$ to the region above this energy.This same quantity($PG_{loss}$) also provides a measure of the loss of metallicity due to the pseudogap in as much as it can usefully be characterized by a single number. Applying the method to the data of Dai {\it et al.}\cite{dai:2012} on the Bi-2201 family, provides a boson spectrum which is very similar in its main characteristic with that found before for Hg-1201\cite{yang:2009} and Bi-2212\cite{hwang:2007}. These previous studies were restricted to overdoped and optimally doped samples as no pseudogap was included in the analysis. Here we find that the trends with doping established previously, not only apply to Bi-2201 as well, but also continue even into the highly underdoped regime. All spectra for Bi-2201 have a pronounced peak at a frequency $\Omega_{sf}$ which tends to decreases with decreasing doping ($p$). This peak is superimposed on a background which extends to very high energy as was observed in Bi-2212\cite{hwang:2007} and Hg-1201\cite{yang:2009} and confirmed in Ref. \cite{hwang:2012}. A valley forms between peak and background as doping is reduced.

An important conclusion of our work is that the quasiparticle mass enhancement factor given as twice the first inverse moment of the electron-boson spectral density $I^2\chi(\Omega)$ continues to increase with decreasing doping $p$ even in the most underdoped sample studied. This trend can easily be understood if we assume that the coupling to the charge carriers is dominantly due to exchange of spin fluctuations which becomes larger as the antiferromagnetic phase is more closely approached. At the same time the pseudogap becomes larger and the material loses metallicity as the Mott insulating state is approached.

\ack

JH acknowledges financial support from the National Research Foundation of Korea (NRFK grant No. 20100008552). JPC was supported by the Natural Science and Engineering Research Council of Canada (NSERC) and the Canadian Institute for Advanced Research (CIFAR).

\section*{References}
\bibliographystyle{unsrt}
\bibliography{bib}

\end{document}